\documentclass[journal=ancac3,manuscript=letter]{achemso}

\usepackage[version=3]{mhchem} 
\usepackage{achemso,array,booktabs,lmodern}
\usepackage[osf]{mathpazo}
\usepackage[scaled=0.95]{helvet}
\usepackage[final]{listings,microtype}
\usepackage[T1]{fontenc}
\usepackage[numbered]{hypdoc}
\usepackage{hyperref}

\DeclareUnicodeCharacter{2212}{-}
\DeclareUnicodeCharacter{2061}{}
\author{Fran\c{c}ois Joint}
\affiliation[CNAM]
{Center for Nanophysics and Advanced Materials, University of Maryland, College Park, MD 20740, USA}
\altaffiliation{Current address: Chalmers University of Technology, Sweden}
\email{joint@chalmers.se}
\author{Kunyi Zhang}
\affiliation[IREAP]
{Institute for Research in Electronics and Applied Physics, University of Maryland, College Park, Maryland
20742, USA}
\author{Jayaprakash Poojali}
\affiliation[University of Maryland]
{Center for Nanophysics and Advanced Materials, University of Maryland, College Park, MD 20740, USA}
\author{Daniel Lewis}
\affiliation[University of Maryland]
{Department of Electrical and Computer Engineering University of Maryland, College Park, MD 20740, USA}
 \alsoaffiliation[IREAP]
{Institute for Research in Electronics and Applied Physics, University of Maryland, College Park, Maryland
20742, USA}
\author{Michael Pedowitz}
\affiliation[University of Maryland]
{Department of Electrical and Computer Engineering University of Maryland, College Park, MD 20740, USA}
\alsoaffiliation[IREAP]
{Institute for Research in Electronics and Applied Physics, University of Maryland, College Park, Maryland
20742, USA}
\author{Brendan Jordan}
\affiliation[University of Maryland]
{Department of Electrical and Computer Engineering University of Maryland, College Park, MD 20740, USA}
\alsoaffiliation[IREAP]
{Institute for Research in Electronics and Applied Physics, University of Maryland, College Park, Maryland
20742, USA}
\author{Gyan Prakash}
\affiliation[IREAP]
{Institute for Research in Electronics and Applied Physics, University of Maryland, College Park, Maryland
20742, USA}
\author{Ashraf Ali}
\affiliation[Univeristy of Maryland]
{Department of Physics, University of Maryland, College Park, MD 20742}
\author{Kevin Daniels}
\affiliation[University of Maryland]
{Department of Electrical and Computer Engineering University of Maryland, College Park, MD 20740, USA}
\alsoaffiliation[IREAP]
{Institute for Research in Electronics and Applied Physics, University of Maryland, College Park, Maryland
20742, USA}
\author{Rachael L. Myers-Ward}
\affiliation[NRL]
{US Naval Research Laboratory, Washington, DC 20375, USA}
\author{Thomas E. Murphy}
\affiliation[Univeristy of Maryland, IREAP]
{Institute for Research in Electronics and Applied Physics, University of Maryland, College Park, Maryland
20742, USA}
\author{Howard D. Drew}
\affiliation[University of Maryland]
{Center for Nanophysics and Advanced Materials, University of Maryland, College Park, MD 20740, USA}
\email{hdrew@umd.edu}

\title[An \textsf{achemso} demo]
  {Terahertz Antenna Impedance Matched to a Graphene Photodetector}

\keywords{THz sensor, graphene, antenna, FIR}

\begin{document}


\begin{abstract}
 Developing low-power, high-sensitivity photodetectors for the terahertz (THz) band that operate at room temperature is an important challenge in optoelectronics. In this study, we introduce a photo-thermal-electric (PTE) effect detector based on quasi-free standing bilayer graphene (BLG) on a silicon carbide (SiC) substrate, designed for the THz frequency range. Our detector's performance hinges on a quasi-optical coupling scheme, which integrates an aspherical silicon lens, to optimize impedance matching between the THz antenna and the graphene p-n junction. At room temperature, we achieved a noise equivalent power (NEP) of less than 300 $pW/\sqrt{Hz}$. Through an impedance matching analysis, we coupled a planar antenna with a graphene p-n junction, inserted in parallel to the nano-gap of the antenna, via two coupling capacitors. By adjusting the capacitors and the antenna arm length, we tailored the antenna’s maximum infrared power absorption to specific frequencies. The sensitivity, spectral properties, and scalability of our material make it an ideal candidate for future development of far-infrared detectors operating at room temperature.
\end{abstract}

\section{Introduction}
§Far-infrared detectors and heterodyne mixers that operate at room temperature are crucial for the deployment on resource-limited platforms such as observation satellites\cite{wiedner2018}, balloons, and small-satellites. These components are pivotal in advancing astrophysical and planetary heterodyne receivers. The main elements of optical receivers—the detector, the local oscillator (LO)\cite{joint2019}, the first amplifier\cite{yebesHEMT}, and the back-end electronics\cite{belitsky2018alma}—must all be state-of-the-art. Space missions, in particular, require components that are low in mass and volume and can operate within the constraints of limited cooling power (typically $\leq$ 100 mW at 4 K) and overall mission power. Therefore, sensors and instrumentation that dissipate low power and can operate at higher temperatures are desirable.

Graphene, known for its high room-temperature carrier mobility, substantial current flux, and high saturation velocity, emerges as an excellent candidate for low-power dissipation and consumption in terahertz (THz) electronics. Various detection mechanisms in graphene photodetectors for the THz frequency range have been explored, including the bolometric effect\cite{yan2012, miaoGrapheneBasedTerahertzHot2018, lara-avila2019}, photo-thermoelectric effect\cite{cai2014, castillaFastSensitiveTerahertz2019, asgariChipScalableRoomTemperatureZeroBias2021, skoblin_graphene_2018}, Dyakonov-Shur (DS) rectification\cite{spirito2014, bandurin2018, bandurin2018a}, ballistic rectification\cite{auton2017, hemmetter2021}, and thermopile\cite{hsu_graphene-based_2015}. At cryogenic temperatures, the hot-electron effect in graphene is well established, where strong electron-electron interactions lead to significant temperature dependence in graphene's conductivity $\sigma$, thereby affecting the resistance R(T) and supporting the development of low-temperature bolometers\cite{miaoGrapheneBasedTerahertzHot2018, lara-avila2019}. Conversely, at room temperature, the hot-carrier-assisted photo-thermoelectric (PTE) effect is notably efficient due to effective carrier heating and significant electronic temperature gradients ($\Delta T_{e}\approx 1000K$\cite{miaoGrapheneBasedTerahertzHot2018}), enhanced by the high Seebeck coefficient ($S\approx100\,\mu V/K$\cite{hwang2009,nam2010, zuev2009}).

The efficacy of these detection systems heavily depends on the quasi-optical coupling efficiency between the incident THz field and the graphene detector. When integrated with planar antennas, the impedance mismatch between the antenna and the graphene sensing region could critically impair the performance of THz detection systems. Various antenna integrated graphene-based photodetectors have been reported, utilizing configurations such as dipole\cite{castillaFastSensitiveTerahertz2019,asgariChipScalableRoomTemperatureZeroBias2021}, broadband log-spiral\cite{asgariChipScalableRoomTemperatureZeroBias2021, bandurin2018, skoblinGrapheneBolometerThermoelectric2018}, and split-bowtie antennas\cite{zak2014}. These detectors use graphene grown via chemical vapor deposition (CVD) on a substrate of 300 nm $SiO_2$ over low-doped Si (100-250 $\Omega.cm$), achieving external voltage responsivities ($\mathbb{R} = V_{ph}/P_{in}$) of up to $\approx15 V/W$ at zero bias voltage with a split bow-tie antenna architecture. Enhancing $\mathbb{R}$ necessitates antennas with a large effective aperture area and correspondingly high gain, and a flat impedance over a wide range of frequencies is desirable. Therefore, optimizing the interface between the graphene detector and the antenna is crucial to minimize coupling losses, and particular attention must be paid to improving the impedance matching between these components. Conventional configurations connect the graphene sensor electrically to the two poles of the antenna\cite{miaoGrapheneBasedTerahertzHot2018, castillaFastSensitiveTerahertz2019, spirito2014}, but their performance rapidly deteriorates at very high frequencies due to impedance mismatch, while the detector's spectral properties remain narrow band\cite{lucas2018, spirito2014, qinRoomtemperatureLowimpedanceHighsensitivity2017}. Overcoming these challenges, such as by using high mobility hBN encapsulated SLG or BLG and optimizing contact resistance, could reduce the limiting effects of thermal noise in graphene detectors.

According to the Wiedemann–Franz law and the Mott relation, the electron thermal conductivity ($\kappa$) of graphene can be expressed as $\kappa = L\sigma T$, where $\sigma$ is the conductivity and $L$, the Lorentz number, is defined as $L = (\pi^2 k_B^2)/(3e^2)$. The Seebeck coefficient ($S$) is given by $S = LT (\partial \ln \sigma / \partial E_F)$, linking it directly to changes in conductivity with respect to the Fermi energy ($E_F$). When a thermal difference ($\Delta T_e$) is induced, it results in a voltage $V = -S \Delta T_e$ and a heat flux to the substrate $Q = G_{th}A\Delta T_{e}$.

For conditions where $T \ll T_F = E_F/k_B$, the responsivity $\mathbb{R} = V/Q$ simplifies to approximately $\mathbb{R} = 2/(\sigma E_F)$. Hence, maximizing $\mathbb{R}$ for the photodetector requires minimizing both $\sigma$ and $E_F$. With residual doping levels of epitaxial graphene on SiC near the Dirac point achieving $n_0 \leq 10^{10} \text{cm}^{-2}$\cite{lara-avila2019}, this suggests a potential maximum responsivity of $\mathbb{R} \simeq 10^3 \text{V/W}$.

Increasing the $T_e$ gradient through enhanced optical absorption can locally augment heat absorption in the BLG\cite{cai2014}. This enhancement was facilitated by integrating subwavelength nano-gap confinement and boosting the near-field light−matter interaction, previously demonstrated to elevate responsivity\cite{castillaFastSensitiveTerahertz2019,egglestonOpticalAntennaEnhanced2015,bettenhausen_impedance_2019,miseikisUltrafastZeroBiasGraphene2020}. Notably, $\mathbb{R}$ is inversely proportional to the thermal conductance ($G_{th}$); thus, a low $G_{th}$ is critical for high sensitivity in THz detection. While epitaxial single-layer graphene (SLG) has shown $G_{th} \simeq 10^4 \text{W/K.m}^2$ at 6K\cite{elfatimy2016}, significantly lower than SLG on $SiO_2$\cite{mak}, our quasi-free standing BLG could potentially exhibit similar or even lower $G_{th}$.

Using a simplified model of PTE detector operation with uniform carrier density across the p-n junction, we anticipate $\mathbb{R} = 2S⁄(A \cdot G_{th}) \simeq 30 \text{V/W}$, given $G_{th} = 7 \times 10^4 \text{W/K.m}^2$\cite{castillaFastSensitiveTerahertz2019}, a Seebeck coefficient of $S = 50 \mu V/K$, and dimensions $L = 4\mu m$ and $W = 35\mu m$. Because $S \rightarrow 0$ at $T = 0$, the PTE effect diminishes at low temperatures, making graphene an attractive material for room temperature operation of THz detectors, potentially rivaling or exceeding existing technologies\cite{liuRecentProgressDevelopment2021}.

In this study, we present a sensitive photo-thermal-electric (PTE)-based antenna integrated graphene photodetector (Figure \ref{fig1}) optimized for the sub-millimeter wavelength, achieving a voltage responsivity ($\mathbb{R}$) of approximately 35 V/W at zero source-drain bias. This system features a fractional bandwidth of about 150 GHz and a noise equivalent power (NEP) at room temperature of 300 $pW/\sqrt{Hz}$. The design was refined through advanced microwave simulations and equivalent circuit modeling, leading to the implementation of a capacitively coupled dipole antenna system. We investigated two antenna configurations: narrow and wide dipole antennas, the latter hereafter referred to as the 'patch' antenna. Our analysis indicated that the dimensions of the antenna strips and the coupling capacitance significantly influence detection magnitude, resonance frequency, and operational bandwidth.

The gapped nature of BLG\cite{kim2013coexisting, ohta2006controlling} plays a pivotal role in enhancing the device's sensitivity through the Seebeck effect, which behaves differently at the p-n junction compared to single-layer graphene (SLG). In BLG, the presence of a bandgap at the charge neutrality point (CNP) of approximately $\Delta \approx 250$ meV facilitates effective modulation of carrier density, enhancing THz frequency responsivity\cite{spirito2014} through differential Seebeck effects across the p-n junction. The maximum Seebeck coefficient difference on both sides of the junction, which scales linearly with the bandgap, arises when the Fermi level crosses the conduction band in the n-region and the valence band in the p-region, boosting the photovoltage responsivity \cite{titova2023ultralow}. The presence of a bandgap not only enhances the Seebeck coefficient but also reduces the phonon-aided cooling due to a decreased density of available states for interband electron transitions, which diminishes completely when $\hbar\omega_{ph}=\Delta$. This reduction in phonon-aided cooling, coupled with a decrease in electronic heat conductance, leads to stronger localized electron heating ($\Delta T_e$) hence elevating the device's responsivity ($\mathbb{R}$).

Our device utilizes quasi-free standing bilayer graphene (BLG), produced via thermal decomposition of silicon carbide and hydrogen intercalation \cite{emery2014structural} on a silicon-carbide substrate. This method not only ensures potentially ultra-low thermal conductance but also high transparency in the THz frequency range, yielding wafer-scale graphene with enhanced carrier mobility and density. These conditions are crucial for achieving low overall resistance, vital for effective impedance matching between graphene and the receiving antenna.

We opted for BLG over single-layer graphene (SLG) due to the more effective modulation of carrier density in BLG, which enhances THz frequency responsivity\cite{spirito2014}. The design incorporates Au split-gates to electrostatically create a p−n junction within the BLG channel and to optimize THz signal coupling onto the BLG. By exploiting optical field enhancement and confinement in the antenna/split gate nano-gap, alongside impedance matching, we significantly enhance the interaction and optical absorption in the p−n junction region. This results in a confined electron heat source and elevated responsivity ($\mathbb{R}$), paving the way for high-performance THz detectors.

\section{Results and discussion}
\subsection{Design and Simulations}

\begin{figure}[htp]
\includegraphics{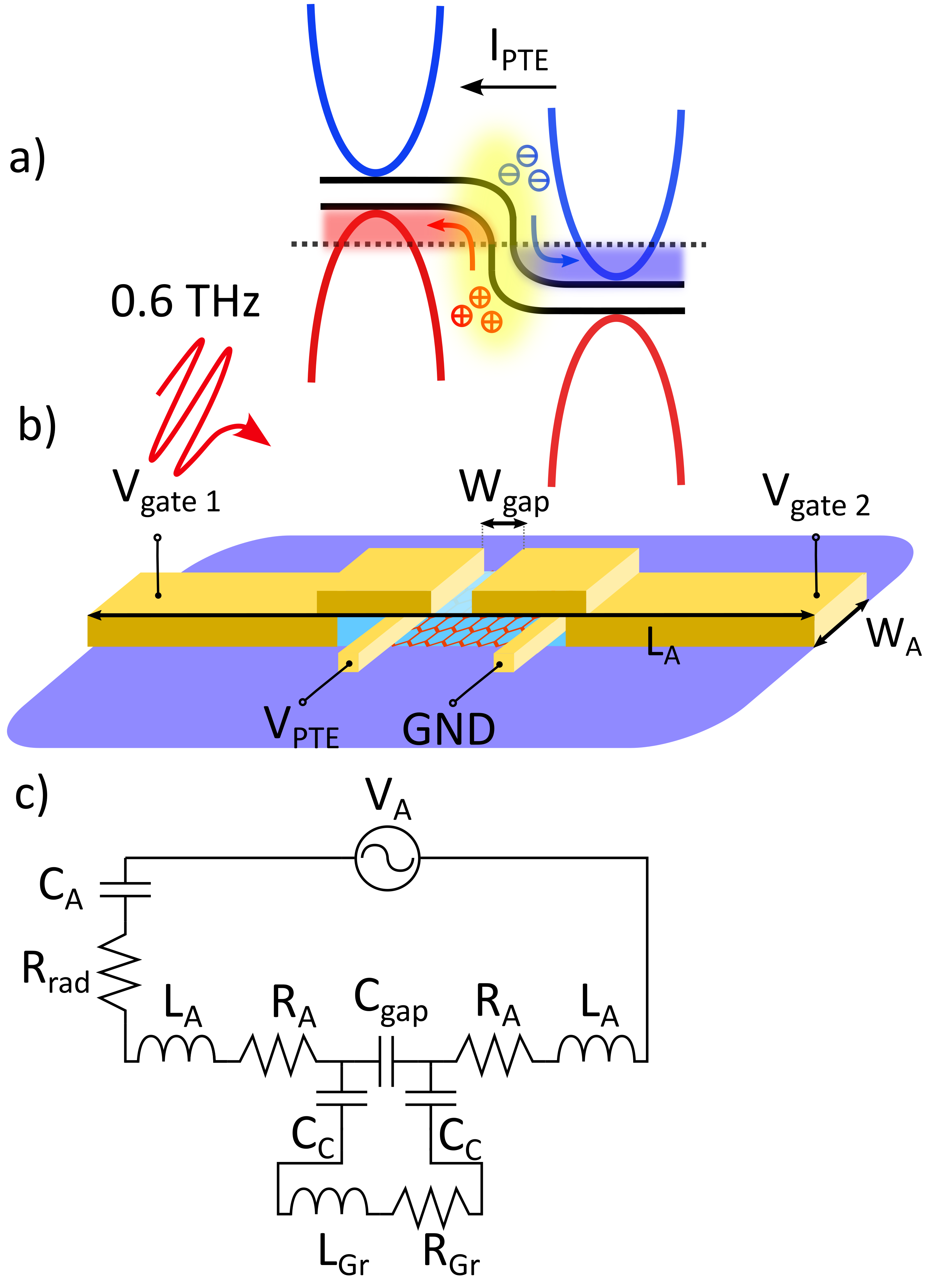}
\caption{ \textbf{a} THz dipole antenna utilizing the PTE effect in BLG as a rectification mechanism. \textbf{b} presents a schematic illustrating the electrical connections to the antenna, where the photovoltage ($V_{PTE}$) is developed between the two antenna arms with the THz field concentrated at the antenna nanogap. Voltages $V_{gate 1,2}$ control the carrier densities on both sides of the gap. \textbf{c} shows the equivalent circuit model for the antenna.
}
    \label{fig1}
\end{figure}

\begin{figure}[htp]
    \centering
\includegraphics{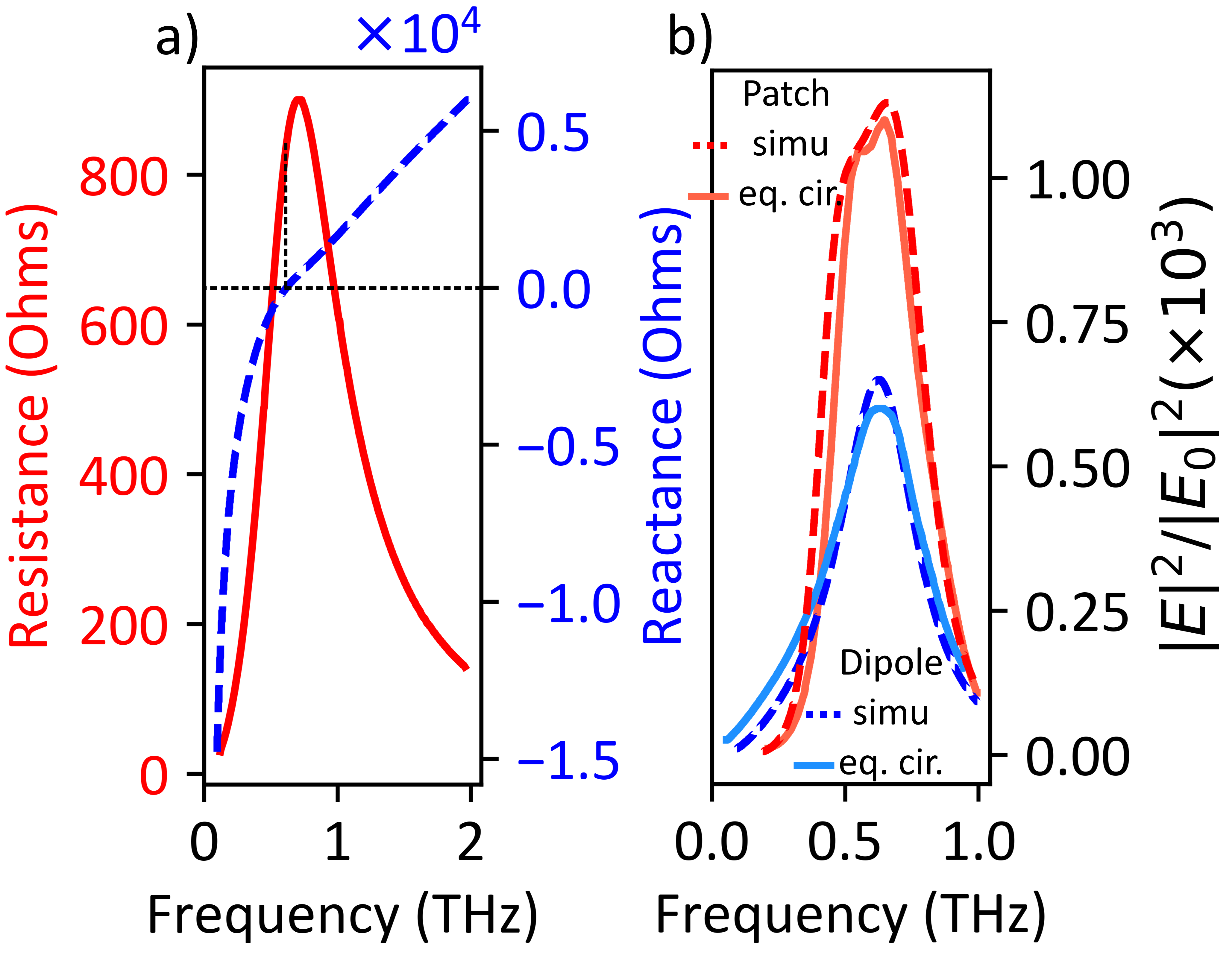}
\caption{\textbf{a} Input resistance and reactance of the antenna are calculated from the equivalent circuit model when matched to a BLG p-n junction at 600GHz. In the impedance matching condition, the antenna's maximum absorbed power occurs near the resonance frequency, which is also verified using FEM simulations, plotted concurrently in \textbf{b}.}
    \label{fig2}
\end{figure}

To optimize the impedance matching for $f=600\,GHz$, we build a lumped element model\cite{bettenhausen_impedance_2019} to calculate the input impedance of the THz antenna and estimate the maximum absorbed power in the BLG based on finite element method (FEM) simulations. The model presented in Figure \ref{fig1}c resembles the shape of the real dipole antenna. Showing all the components modeled and described here: the model accounts for the ohmic losses in the antenna’s arm and the radiation losses ($R_{ohm}, R_{rad}$), the self-capacitance $\scriptstyle C_{A}=\varepsilon_{0}L_{A}/(ln\frac{L_{A}}{W_{A}})$ as well as the kinetic inductance of the gold ribbons $L_{k}=\frac{L_{A}}{A}Re(\frac{1}{\omega^{2}\varepsilon_{0}(1-\varepsilon_{Au})})$ \cite{kraus1988antennas}. A gap capacitance $C_{gap}=\frac{\varepsilon_{0}A}{d}$ was introduced for coupling the antenna arms at the nano-gap. The coupling capacitors  $C_C$ represent the capacitive coupling between the antenna’s arm and the BLG, and modeled as an additional sheet inductor. The source of the two circuits is defined by the open-circuit voltage source \cite{bettenhausen_impedance_2019} $V_{A}$, which is equal to the projection of the incident field intensity $E_{0}$ in $V/m$ to the length of the antenna $L_{A}$ ($V_{A}=E_{0}\times L_{A}$). From the equivalent circuit in Figure \ref{fig1}c, we define an expression of the field enhancement at the nano-gap $\vert E_{gap}\vert^{2}/\vert E_{0}\vert^{2}=(\frac{l_{A}}{d_{gap}})^2\vert\frac{Z_{gap}}{Z_{gap}+Z_{A}}\vert^{2}$, where the enhancement is proportional to $1/d^2$. From this model, it is clear that a small antenna gap is desired. The field at the center of the nano-gap $V_{A}$ reaches a maximum when the reactance of the equivalent impedance at the nano-gap $Z_{gap}$ matches the dipole antenna reactance $X_{A}=-X_{gap}$. The gap impedance $Z_{gap}$ comprises the graphene complex impedance $Z_{G}$. The ohmic contacts of the source and drain electrodes to the graphene sheet, $R_{C}$, and the capacitance between the antenna arms with the graphene layer, $C_{C}$ are included in $Z_{gap}$. The first antenna resonance is set by $L_A$ and $C_A$ with $f_A=1/2\pi\sqrt{ L_AC_A}$. $f_A\simeq\,590\,GHz$ for the dipole antenna and $\simeq\,570\,GHz$ for the patch. The antenna operates in an open circuit at the second resonance, as $X_{gap}$ creates a minimum of the sum of gap and antenna impedances. This is shown in Figures \ref{fig2}a,b where the maximum gap field intensity is correlated to the maximum input resistance. The second resonance frequency for both dipole and patch occur at 610 GHz and 625 GHz respectively, merging with the first antenna resonance. Although it is possible to tune the second resonance frequency by changing the complex impedance of the graphene load\citep{bettenhausen_impedance_2019}. FEM simulations were conducted in parallel to confirm the equivalent circuit model. The absorbed power density in the BLG is computed (Figure \ref{fig2}b) at a wide frequency range where we extract the antenna enhancement factor at the nano-gap.

 As PTE detectors are sensitive in the region where the graphene's chemical potential is tuned near the charge neutrality point\cite{asgariChipScalableRoomTemperatureZeroBias2021}, the impedance seen by the antenna is usually much higher than for the ungated graphene channel. To simulate the impedance of the THz detector near the optimal operational point, we use a resistance sheet model\cite{yang2022resistive} with an effective layer thickness $d_{eff}\,=\,2\,nm$ and impedance at 600 GHz  $Z_{sheet}(E_F\,=\,200\,meV)\sim\,(1k\,+\,j500)\,\Omega/sq$, ensuring that the antennas are designed to operate in a regime where the PTE detector is the most sensitive. Both 3D full wave simulation and equivalent circuit model converge to a similar antenna enhancement profile for the dipole and patch antenna. From the FEM  simulations, we can extract the antennas area's effective area $A_{e}$ defined as $A_{e}=P_{gr}/p_{in}$ (in $m^2)$, where $P_{gr}$ and $p_{in}$ are the power absorbed in the graphene load and the incident THz power density (in $W/m^2$), respectively. Using both equivalent circuit models and FEM simulations, the optimised geometrical parameters for the dipole and patch antennas were extracted and summarized in table \ref{tbl0}.

\begin{table}
\begin{center}
  \caption{Antenna design physical parameters.}
  \label{tbl0}
  \begin{tabular}{|p{2cm}||p{2cm}|p{2cm}|p{2cm}|p{1.5cm}||p{1.5cm}|}
  \hline
  & Antenna Dimensions&Channel Dimension&Resonance Freq.&BW&$A_e/\lambda^2$\\
     &$\mu m\times \mu m$&$\mu m\,\times\,\mu m$&GHz&GHz&\\
     \hline
     \hline
     dipole&260$\times$5&5$\times$5&610&320&0.23\\
     \hline
     Patch&180$\times$36&4.5$\times$36&625&450&0.31\\
    \hline
  \end{tabular}
  \end{center}
\end{table}

\subsection{ d.c. Characterisations}
\begin{figure}[htp]
    \centering
\includegraphics[width=\textwidth]{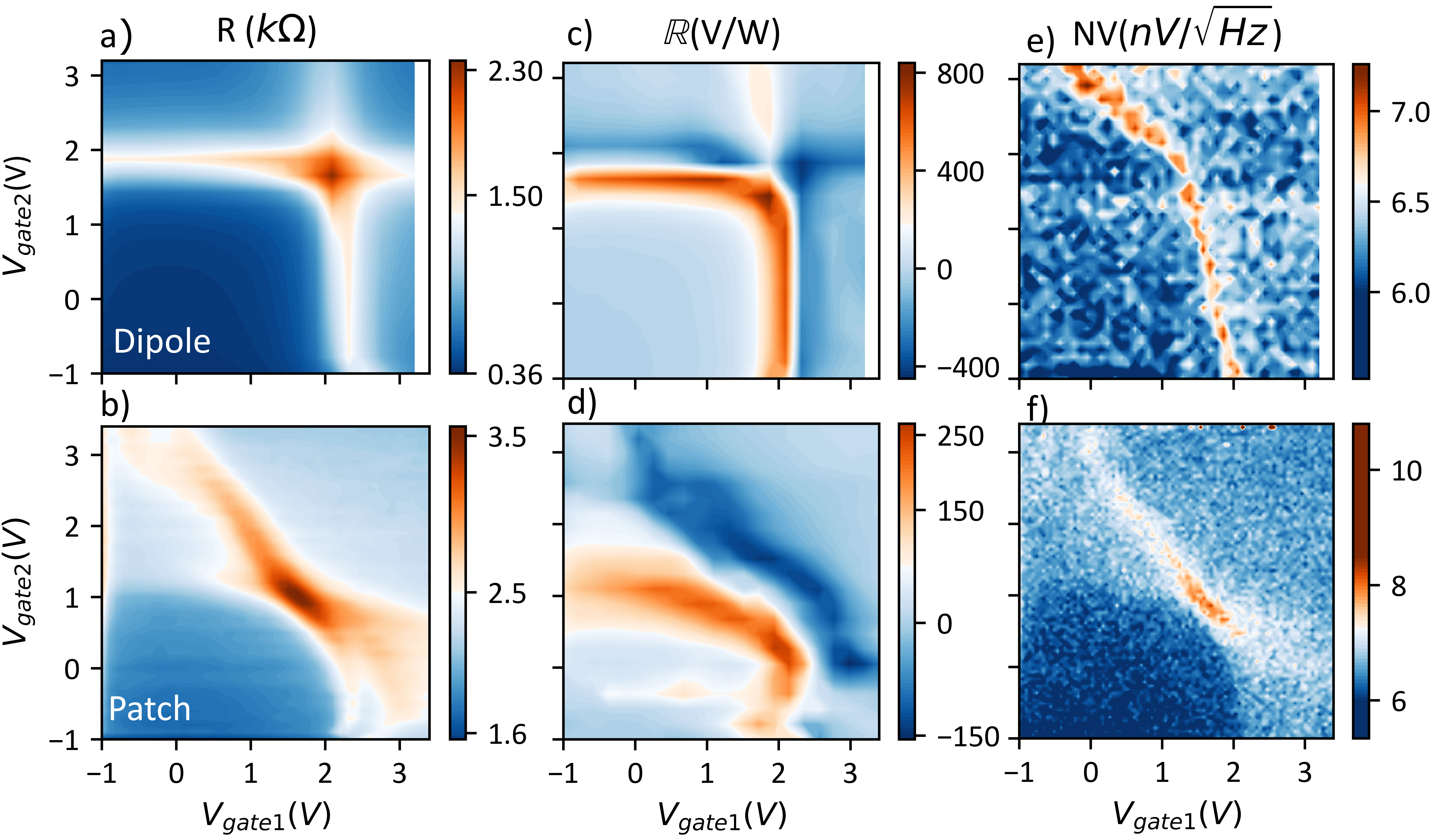}
\caption{Transport characterizations of dipole antenna ( \textbf{a,c,e}) and patch (\textbf{b,d,f}) PTE detectors at room temperature. \textbf{a} and \textbf{b} show resistance maps as a function of the voltage applied to two top gates, revealing four distinct regions corresponding to different doping configurations: p-n, p-p, n-p, and n-n. \textbf{c} and \textbf{d} depict direct current (d.c.) transport responsivity, measured in volts per watt (V/W), against gate voltage without THz excitation. \textbf{e} and \textbf{f} present a noise voltage map as a function of the gate voltage, with no terahertz excitation present.}
    \label{fig3}
\end{figure}

To establish the optimal operating conditions for our photodetector, we conducted d.c. electrical characterizations. This involved sweeping the split-gate voltages ($V_{\text{gate 1}}$, $V_{\text{gate 2}}$) while recording the device current $I_{\text{DS}}$ and varying the source-drain bias $V_{\text{DS}}$ from -5V to 5V. The resistance map of a dipole (patch) antenna p-n junction detector (Figure \ref{fig3}a and b respectively) displays four regions, each corresponding to different doping levels on either side of the junction. The map is symmetric, with resistance peaking at approximately $R\approx3.5k\,\Omega$ for the patch antenna and $R\approx2.2k\,\Omega$ for the dipole at the charge neutrality point (CNP), which ranges between 1.8V and 2.2V. This reflects p-doping of the unbiased bilayer graphene (BLG) with a carrier concentration of $n\approx 1.7 \times 10^{13}\,cm^{-2}$ and a Fermi energy of $E_{F}\approx240\,\text{meV}$ relative to the CNP.

The total resistance $R$ consists of channel resistance ($R_{\text{ch}}$) and contact resistance ($R_{\text{C}}$). $R_{\text{ch}}$ includes a fixed contribution from ungated BLG regions and a gate-dependent contribution from channel segments beneath the split gates. The observed gate-dependent variability in $R$ in Figure \ref{fig3}a and b indicates that $R_{\text{ch}}$ is the predominant factor, aligning with our low contact resistivity (<40 $\Omega.\mu m$) for epitaxial BLG, based on channel geometry and sheet resistance measurements from independent reference samples.

To evaluate the performance of the graphene thermoelectric detector, we conducted d.c. measurements to determine the Joule heating responsivity (Figure \ref{fig3}c and d). A d.c. bias voltage was applied to heat the sample, and the thermoelectric current was measured by comparing electric currents under forward and reverse biases. The peak responsivity in $\mathbb{R}_{d.c.}$ occurs at low carrier densities, where Joule heating is maximal and the Seebeck coefficient is high. The measured noise voltage (NV) with no terahertz excitation for both detector types is shown in Figure \ref{fig3}e and f. NV arises from intrinsic device noise and amplifier noise, which are uncorrelated and combined in quadrature. Within the response bandwidth, intrinsic device noise is primarily due to thermal fluctuations and Johnson-Nyquist noise. The thermal fluctuation noise is given by $NV_{\text{th}}=\sqrt{4\,k_B T^2 G_{\text{th}}}$, where T is the average bridge temperature, estimated at $22\,pV/\sqrt{Hz}$ for a $G_{th}$ of $10^{4}\, W\,K^{-1}\,m^{-2}$. The Johnson-Nyquist noise contribution is $NV_{\text{JN}}=\sqrt{4k_{B}TR}$, calculated at $7.5\,nV/\sqrt{Hz}$ at the CNP ($R=3.6\,k\Omega$) for the patch antenna and $6.0\,nV/\sqrt{Hz}$ ($R=2.2\,k\Omega$) for the dipole antenna. Experimental noise closely matches theoretical predictions, suggesting that detector performance is Johnson-Nyquist noise-limited. From $\mathbb{R}_{d.c.}$ and measured NV, we derive an electrical NEP of $30\,pW/\sqrt{Hz}$ for the patch antenna and $7.5\,pW/\sqrt{Hz}$ for the dipole antenna, comparable to the sensitivities of conventional bolometers\cite{zhang2019fast}.

\subsection{Optical and Spectral Characterisations}
To record the terahertz (THz) responsivity of our photodetector, we aligned a continuous-wave (CW), transverse-electric (TE) polarized THz source with an aspherical lens. The photo-voltage ($V_{\text{PTE}}$) was measured between the source and drain electrodes across the unbiased channel ($V_{\text{DS}}=0$V). Measurements were taken by varying the split-gate voltages ($V_{\text{gate 1}}$, $V_{\text{gate 2}}$) and employing a lock-in amplifier with internal modulation (square wave, ON-OFF) of the THz source at 1 kHz. The largest photo-response occurred in the central p-n and n-p junction regimes close to the Dirac point, where the antenna's nanogap optimally channels the THz radiation.

\begin{figure}[htp]
    \centering
\includegraphics[width=\textwidth]{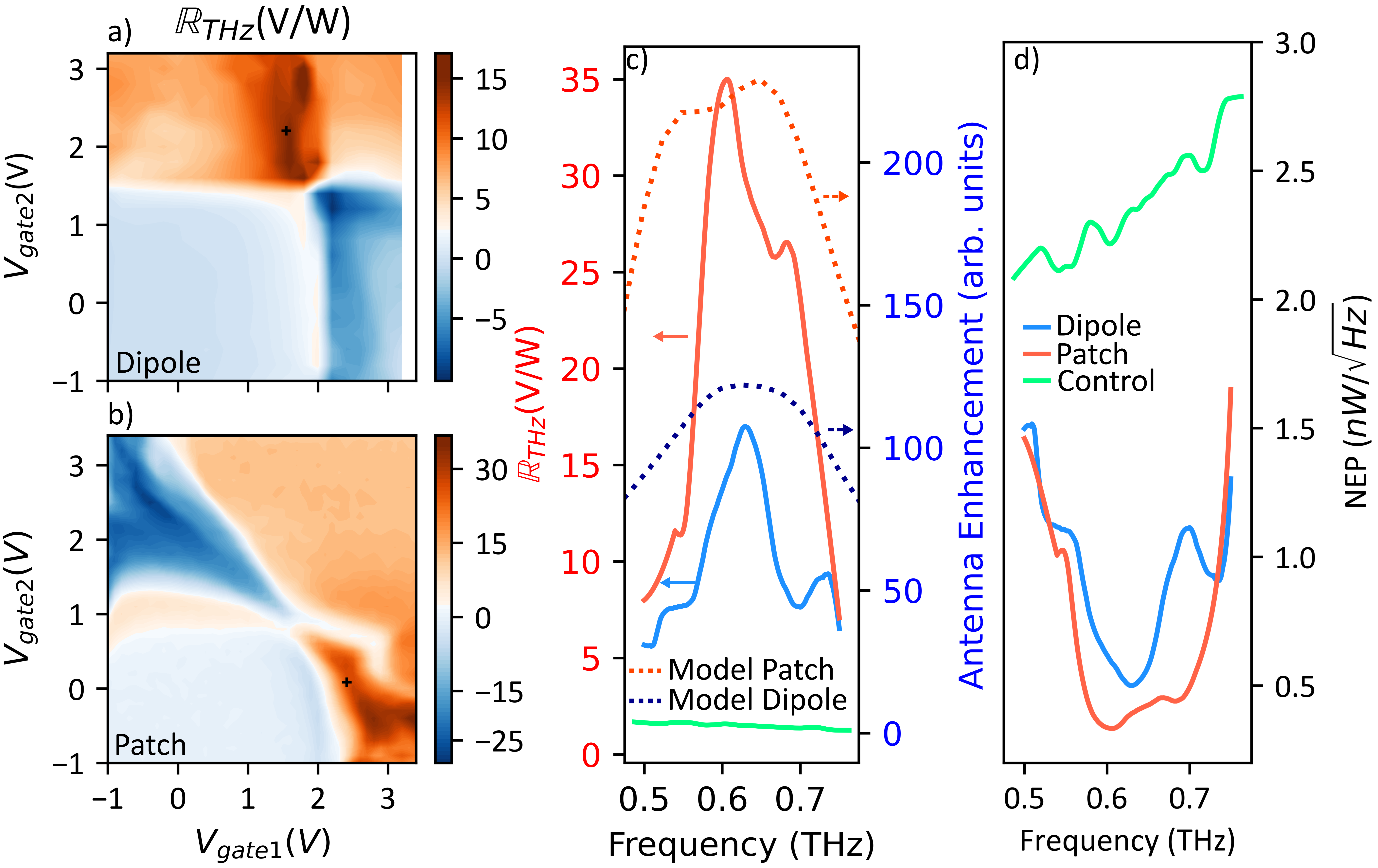}
\caption{ Responsivity maps and NEP spectra for patch and dipole antennas demonstrating PTE. \textbf{a} and \textbf{b} show THz responsivity maps for the patch and dipole antennas, respectively, each featuring a six-fold symmetry characteristic of the photo-thermal electric effect. \textbf{c} displays the antenna-enhanced responsivity spectra. For both NEP and $\mathbb{R}_{THz}$ spectra, the used set of gate voltages is indicated in the responsivity maps in \textbf{a} and \textbf{b} with circles. \textbf{d} illustrates the antenna-optimised NEP in $W/Hz^{1/2}$ for both dipole and patch antenna detectors, measured at optimal gate voltage configurations. Concurrent plots of the calculated antenna enhancement from the circuit model are provided for both types of antennas, alongside the NEP and responsivity of an unmatched antenna PTE detector (control).
}
    \label{fig4}
\end{figure}

To compute the THz responsivity ($\mathbb{R}_{\text{THz}}$), we estimated the optical power received by the bilayer graphene (BLG) based on the power intensity at the lens focus spot, defined as $Intensity = P_{\text{in}} / S_{\text{A}}$, where $P_{\text{in}}$ is the total measured incident power and $S_{\text{A}}$ is the lens focus spot area. The absorbed power by the BLG, $P_{\text{abs}}$, is then calculated as $Intensity \times A_{\text{e}}$, with $A_{\text{e}}$ being the effective aperture area determined through finite element method (FEM) simulations (Table \ref{tbl1}). We observed maximum responsivities of about 15 V/W for the narrow dipole antenna and 35 V/W for the patch antenna (Figure \ref{fig4}a,b).

The shape and magnitude of both d.c. transport and optical responsivity indicate that signals originate from distinct p-n junctions within the device. Remarkably, the magnitude of $\mathbb{R}_{\text{d.c.}}$ exceeds $\mathbb{R}_{\text{THz}}$ by more than an order of magnitude, a trend also noted for the narrow dipole antenna. While we have observed significant improvements, there remains considerable potential for optimizing the quasi-optical coupling between the antenna and the lens. The response spectra of both the dipole and patch antennas are displayed in Figure \ref{fig4}c, with gate voltages optimized to maximize photovoltage responses at the p-n junctions (black cross in the photo-voltage maps). The presence of ripples ($f_{\text{ripples}}$= 5 GHz), potentially due to the uncoated high-resistivity silicon (HR-Si) lens affecting power coupling efficiency, is noted in the measurements (\hyperref[supp]{Supporting Information}).

Filtered spectra reveal a clear enhancement of responsivity at the antenna's resonance frequency, aligning well with the targeted frequency of 600 GHz. The spectrum's dependence of the optimization of gate voltages near the CNP, where $\mathbb{R}_{\text{THz}}$ peaks, underscores the critical role of antenna design. The responsivity spectral bandwidth correlates with the $l/d$ ratio \cite{balanis2005}, where $l$ and $d$ are the length and width of the antenna, respectively. For the dipole antenna ($l/d = 26$), the fractional bandwidth $\Delta f/ f_0$, where $\Delta f$ is the $-3dB$ bandwidth of the antenna resonance and $f_0$ the resonance frequency, is about 15.8$\%$, and for the patch antenna ($l/d = 2.6$), it increases to 28.2$\%$

Using the extracted responsivity, we evaluated the sensitivity of the detector against the frequency of the impinging THz field (Figure \ref{fig4}d). By dividing the thermal noise voltage variance per 1 Hz of bandwidth by the responsivity, the noise equivalent power (NEP) was calculated as approximately 300 $pW/\sqrt{Hz}$ for the patch and 400 $pW/\sqrt{Hz}$ for the dipole antenna at their resonant frequencies at 300K (see Figure\hyperref[supp]{3} in \hyperref[supp]{Supporting Information}). Both NEP and responsivity spectrum measurements verify that the spectral ranges where the THz detectors operate are significantly enhanced by the matched antennas compared with a control detector, which has an antenna matched at 2500 GHz, well outside the studied frequency range.

\begin{table}
\begin{center}
  \caption{Comparison of Antenna Coupled Graphene photo-detector and to other room-temperature THz sensors with various detection mechanisms.}
  \label{tbl1}
  \begin{tabular}{|p{2cm}||p{2cm}|p{1.5cm}|p{1.5cm}|p{2cm}|p{2cm}|p{1.0cm}|}
    \hline
    Detection Mechanism  & Resonance Freq.   & $\rm I\!R_{THz, max}$  & $A_e$ Simulation  & Fractional BW & NEP&ref \\
     &(GHz)&(V/W)&$\mu m^2$&(GHz)&$pW/\sqrt{Hz}$&\\
    \hline
    \hline
    PTE & 634  & 15 & $0.23\lambda^2$&$15.8\%$&400& this work\\
    PTE & 600  & 35 & $0.31\lambda^2$&$28.2\%$&300&\\
    \hline
 PTE& 325  & 30 & $0.1\lambda^2$&$4.6\%$&51& \cite{qinRoomtemperatureLowimpedanceHighsensitivity2017}\\
 \hline
 PTE& 2800 & $\approx$6&n.c.&$12.5\%$&$\sim1000$&\cite{asgariChipScalableRoomTemperatureZeroBias2021}\\
 \hline
  PTE & 1800  & 30 & $0.31\lambda^2$&n.c.&51&\cite{castillaFastSensitiveTerahertz2019}\\
  \hline
  PTE & 600  & 15 & n.c.&n.c.&515&\cite{zak2014}\\
  \hline
  \hline
  CMOS & 650  & 70k & n.c.&n.c.&300&\cite{lisauskas2009rational}\\
  \hline
   InGaAs Diodes & 650  & 13 & n.c.&n.c.&200&\cite{palenskis2018ingaas}\\
    \hline
   MEMS bolometer& 250-3000&  & n.c.&n.c.&500&\cite{zhang2019fast}\\
   
  \hline
  \end{tabular}
  \end{center}
\end{table}

In this study, we demonstrated the capabilities of quasi-free standing bilayer graphene (BLG) on silicon carbide (SiC) in developing antenna-enhanced graphene photodetectors. Our detectors achieved an external responsivity of approximately 35 V/W, a noise equivalent power (NEP) of about 300 $pW/Hz^{1/2}$ at 300 K, and a $-3$dB spectral width of approximately 150 GHz. These results were made possible by the integration of a terahertz (THz) antenna, which was impedance-matched to a BLG p-n junction, optimizing the light−BLG interaction and creating a confined electron heat-source that predominately generates a photo-thermal-electric (PTE) signal.

Our modeling of the detector architecture allowed us to maximize the absorption and $\mathbb{R}_{THz}$ at a specified frequency, while also providing the flexibility to adjust the detector's operational bandwidth. The notable improvement in NEP can be attributed to the enhanced responsivity, a direct consequence of our precise antenna design. Moreover, the high $\mathbb{R}_{d.c.}$ observed suggests that quasi-free standing BLG on SiC is a promising material for the future development of scalable THz sensors capable of operating at room temperature. These findings not only underscore the potential of graphene-based devices in THz applications but also open avenues for further innovations in optoelectronic sensing technologies.

\begin{acknowledgement}
The authors gratefully acknowledge the support and facilities provided by the staff of the University of Maryland Fabrication Laboratory. This work was supported by the National Aeronautics and Space Administration under Grant No. 80NSSC18K0933 through the Mission Directorate. Work at the U.S. Naval Research Laboratory is supported by the Office of Naval Research.

\end{acknowledgement}

\begin{suppinfo}
\label{supp}
Experimental methods, electrical and optical characterization of 600 GHz and 2.5 THz dipole devices, optical ray tracing simulation of the HR-Si lens, FEM simulation of the antennas, circuit model derivations, and cryogenic characterizations of the sensors at 4K.
\end{suppinfo}

\bibliography{PTE_Graphene}

\end{document}




\section{Methods}
\subsection{Device Fabrication}
High-quality quasi-free standing epitaxial graphene (QFS EG) is synthesized via Si sublimation followed by hydrogen intercalation on a 4-inch diameter, semi-insulating (0001) $0.1^\circ$ off-axis 6H–SiC wafer, using an Aixtron/Epigress VP508 horizontal hot-wall reactor. After dicing into $8,mm \times 8,mm$ pieces, graphene channels are patterned using electron beam lithography with a 400 nm thick layer of PMMA (poly(methyl methacrylate), Micro Chem Corp.), followed by etching with inductively coupled plasma. The gate dielectric, consisting of $SiO_2$, typically measures 120 nm in thickness and is deposited by electron beam evaporation. The antenna arms are fabricated via electron beam lithography (EBL) and a lift-off process using Cr/Au (5/80 nm). The gaps for the dipolar antennas measure 200 nm and 400 nm for the dipole and patch antennas, respectively. The detector chip is then mounted onto the back of a custom-made, high-resistivity silicon hemi-aspherical lens using cyanoacrylate glue. For all devices presented in this document, the graphene mobility is approximately $3500\,cm^2/V.s$ with a carrier density of $n\approx 1.7 \times 10^{13}\,cm^{-2}$.

 \subsection{d.c. Characterization}
DC measurements were performed by wire bonding the source-drain channel and the two split gates to a read-out PCB board. A resistance map of each device was obtained by recording the I-V characteristics of the channel and sweeping the voltages applied to the split gates. The channel resistance was derived from a linear regression of the I-V curve. The DC heating responsivity shown in Figure 2b was measured using a Keithley 2400 source measure unit to minimize voltage noise. We applied DC voltages of equal magnitude but opposite directions across the source and drain contacts and measured the currents in both cases. When the bias voltage's sign was switched, the bias current reversed direction, while the thermoelectric current maintained its direction. The rectification current (the difference in current under forward and reverse bias) multiplied by the sample resistance yielded the thermoelectric voltage. This voltage was then normalized by the electric heating power, enabling the extraction of thermoelectric voltage responsivity.

Noise Measurements: Noise measurements were carried out using a lock-in amplifier (EGG 7280) set at a reference frequency of 1 kHz. The gain of the lock-in amplifier was excluded from the quoted noise spectral density, the photovoltage, and the voltage responsivity reported.

The DC thermoelectric responsivity was characterized by applying a DC voltage ($\pm V$) across the graphene channel and measuring the resulting currents $I_1 = I + I_{\text{thermal}}$ and $I_2 = I - I_{\text{thermal}}$, where $I$ is the current from the bias voltage, and $I_{\text{thermal}}$ is the thermoelectric current. The DC responsivity, $\mathrm{IR_{d.c.}}$, was then calculated as $\mathrm{IR_{d.c.}} = \frac{I_{\text{thermal}}}{I^2}$ (V/W).

The responsivity, $\rm I!R_{THz}$, was measured at 605 GHz for the patch and 625 GHz for the dipole, defined as the voltage change across the graphene channel per unit of absorbed signal power $P_{abs}$, expressed in V/W. $P_{abs}$ is determined using the antenna capture cross-section $A_e = 0.31\lambda^2$ for the patch antenna and the incident power density $P_{in}$, with $P_{abs} = A_e \times P_{in}$.

\subsection{THz Detection.}
The static photo response was characterized using a tunable terahertz source (VDI-AMC). The terahertz wave, generated from an input microwave signal, has its frequency ($f_{mw}$) tunable from 9.00 GHz to 14.27 GHz. This input is power-modulated into a square-wave pattern at 10 dBm. The output terahertz frequency is an integer multiple of the input microwave frequency, determined by the frequency multipliers within the terahertz source: $f_{THz} = N \times f_{mw}$, where $N = 56$. The total output power of the terahertz source remains constant at less than $300\mu W$ across the frequency range of study.

The photovoltage response to the incident terahertz radiation was measured using a voltage pre-amplifier (SR 560), followed by a lock-in amplifier. Photovoltage maps were obtained by maintaining zero current between the source and drain electrodes while sweeping the voltages applied to the split gates. The modulation frequency of the terahertz source is adjustable from a few tens of Hz to 10 kHz, with a fixed duty cycle of 50$\%$.

\subsection{Circuit Model Derivation}
The capacitively coupled dipole antennas are described using a circuit model composed of lumped elements that account for ohmic losses $R_{\text{ohm}}$, radiation losses $R_{\text{rad}}$ \cite{kraus1988antennas}, and the capacitance and inductance of the antenna arms. The inductance $L_A$ is divided into two components: the Faraday inductance $L_F$ \cite{huang2021antennas, kraus1988antennas}, which is solely dependent on geometrical parameters, and the kinetic inductance $L_{\text{kin}}$ \cite{balanis_antenna_2005}, defined by the imaginary part of the material’s resistivity:
\begin{equation}
Z_{A}=R_{\text{ohm}}+R_{\text{rad}}+j\left(\omega(L_F+L_{\text{kin}}) - \frac{1}{\omega C_A}\right)
\end{equation}
\begin{equation}
L_{\text{kin}}=\text{Im}(\rho(\omega))\frac{l}{A}
\end{equation}

The antenna's capacitance $C_{A}$ is calculated using the fringe field capacitance model \cite{kim2006modeling} and is defined as $C_A=\epsilon_A L_A \times \ln(L_A/W_A)$. An additional capacitance, resulting from the capacitive coupling of the antenna arms at the central nano-gap, is described by a parallel plate capacitance model:
\begin{equation}
C_{\text{gap}}=\varepsilon_0 \varepsilon_{\text{gap}}\frac{A}{d}
\end{equation}

Alongside the capacitive and inductive components, resistive elements in the model represent ohmic and radiation losses. The resistances are defined as follows:
\begin{equation}
R_{\text{ohm}}=\text{Re}(\rho(\omega))\frac{l}{A}
\end{equation}
\begin{equation}
R_{\text{rad}}=\frac{2\pi}{3}Z_0 \left(\frac{l}{\lambda}\right)^2 n_A
\end{equation}

Here, $n_A$ denotes the refractive index of the surrounding medium of the antenna. These elements are integrated into a series oscillating circuit, as depicted in Figure 1a of the main text. The graphene bi-layer is represented in Figure 1a with $R_{Gr}$ and $L_{Gr}$. The loaded input impedance of the antenna with bilayer graphene (BLG) is obtained as follows:
\begin{equation}
Z_{\text{in}}=Z_A \parallel Z_{\text{gap}}=Z_A \parallel (Z_{Gr} \parallel C_{\text{gap}})
\end{equation}

The accompanying table details the equivalent circuit parameters used in the model.

\begin{table}
\begin{center}
  \caption{Equivalent circuit parameters.}
  \label{tbl1}
  \begin{tabular}{|p{2cm}||p{2cm}||p{2cm}|}
    \hline
    &Dipole&Patch\\
    \hline
   $C_{gap}$&0.022 fF&0.08 fF\\
   $L_{A}$&0.24 pH&0.1 pH\\
   $C_{A}$&0.34 fF&0.87 fF\\
   $C_C $&2.66 pF&19.1 pF\\
   $L_{Gr}$&2.65 nH&0.33nH\\
   $R_{Gr}$&1000$\,\Omega$&160$\,\Omega$\\
    \hline
  \end{tabular}
  \end{center}
\end{table}

\subsection{Field Enhancement}
The field enhancement at the gap of the dipole antenna can be calculated\cite{bettenhausen2019impedance}:
\begin{equation}
  \frac{E_{gap}}{E_{0}}=\frac{l_{A}}{d_{gap}}.\frac{Z_{gap}}{Z_{gap}+Z_{A}}\\
  =\frac{l_{A}}{d_{gap}}.\frac{Z_{gap}}{R_{gap}+j.X_{gap}+R_{A}+j.X_{A}}
\label{eqn6}
\end{equation}
where $l_{A}$ and $d_{gap}$ are the effective length of the antenna and the antenna's gap length, respectively. Equation \ref{eqn6} describes the field enhancement at the nano-gap with $E_{gap}$ and $E_{0}$ the electric field at the nano-gap and the incident electric field, respectively. The field enhancement is maximized when the sum of the gap impedance and the antenna impedance is minimized. This minimum is obtained when $X_{A}=-X_{gap}$, when the gap reactance and the antenna reactance are cancelling out. In this condition, the antenna exhibits an open-circuit resonance. By loading the antenna with an appropriate graphene sheet and 2 loading capacitors, the matching condition can be met. 

To compare the circuit model results with the FEM simulations, we calculate for the electric field intensity in the nano-gap and extract the absorbed power in the BLG (Figure 1d. in the main text). The graphene sheet is modeled as a resistance sheet model given in $\Omega/$square which allows for high accuracy and fast computation times. The contact resistance was introduced in the real part of the impedance of the graphene and no special attention has been paid at for the meshing at the contacts. The antenna is placed on a semi-infinite silicon-carbide substrate and is excited by a TE-polarized plane wave, impinging the antenna from the silicon-carbide side. The simulation domain is enclosed in all directions with perfectly matched layer, which ensure that the antenna is only excited by the incoming plane wave.
The calculated input impedance for the dipole and patch antenna and the absorbed power spectra are shown in Figure 1c and Figure 1d respectively. With this model and the results of the full-wave simulations, the functionality of the circuit model can be verified.

\subsection{Quasi-optical Coupling}
An efficient solution to focus the incoming THz signal onto the graphene photodetector is the use of lenses in materials with similar refractive index as the detector’s substrate. We use high resistivity silicon (HR-Si) which as a refractive index of $n_{Si}=3.40$ in the THz frequency range, which is slightly higher than the SiC refractive index of the detector’s substrate ($n_{SiC}=2.90$). Attaching a lens on the antenna substrate will increase the antenna gain on the dielectric side. The detector’s substrate is glued to the flat side of the hemispherical lens using low viscosity cyanoacrylate glue. The cyanoacrylate glue doesn´t have any absorption features in the THz frequency range that we are interest in (400 GHz to 1100 GHz). The dielectric lens optimized design is hyper-hemi-aspherical with an extension length L on the plane side. At the crossing of two different optical or electromagnetic approach, the effect of the lens on the incoming signal can be seen through the ray optic formulation: the THz incoming signal is concentrated/focused on the antenna: it is mandatory that the antenna is carefully placed at the focus point of the lens to get an optimal coupling, a precision of few dozen $\mu m$ is required. Raytracing techniques (using commercially available Zemax software) are used to compute the characteristic of the integrated lens antenna on dielectric as shown in Fig. S3. The optimization of the extension length L, the asphericity is then optimized by simulation in order to obtain the required focus size.
\begin{figure}[htp]
    \centering
\includegraphics[width=\textwidth]{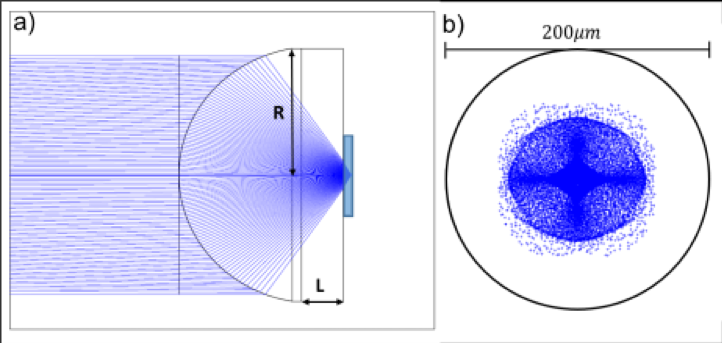}
\caption{a) Profile of the hyper-hemispherical HR-Si lens of radius R and extension length L under consideration with ray-tracing simulation (Zemax). R, L and the asphericity parameter have been optimized with a merit function. The light blue plain rectangle represents the graphene-mixer device in SiC glued on the back of the planar side of the lens. The planar gold patch/dipole antenna lays on the opposite side. The figure S3b) shows the spot image of a circular collimated Gaussian beam impinging on the convex side of the lens on the planar antenna plane. The size of the image spot fits entirely on the antenna effective aperture.
}
    \label{figS3_lens}
\end{figure}
\subsection{Affective Aperture Area}
Another crucial parameter for optimizing the quasi-optical coupling between the impinging THz signal and the graphene detector is the effective aperture ($A_e$) of the dipole receiving antenna. The effective aperture represents the surface area through which power is radiated or received, playing a pivotal role in determining the waist dimension of the incoming Gaussian beam that needs to be coupled to the antenna. The effective aperture of an antenna can be expressed as:

\begin{equation}
A_e = \frac{P_t \text{ (power absorbed in graphene)}}{P_i \text{ (incident power density)}} \label{eqn4}
\end{equation}

Where the incident power density $P_i$ is calculated as $P_i = \frac{E^2}{2\eta_{SiC}}$, with $E=1$ and $\eta_{SiC} = \frac{377}{3.1}$. This results in $P_i = 4.1 \times 10^{-3} \text{ W/m}^2$. The aperture efficiency of an antenna determines the amplitude of power received, which depends on both the energy distribution radiated by the source and the beam pattern of the receiving antenna. For an extended energy source, and particularly when the angular dispersion of the beam exceeds the main lobe of the antenna, the coupling efficiency of the incoming beam to the antenna must be carefully optimized. In this study, due to limitations in computational resources and for simplification, the effects of the lens were omitted in the Finite Element Method (FEM) simulations.
\begin{table}
\begin{center}
  \caption{FEM Simulations results for a narrow dipole antenna ($W_{A}\,=\,5\,\mu m$) and $Z_{Gr}$ taken at $E_F\,=\,300\,meV$}
  \label{tbl1}
  \begin{tabular}{|p{2cm}||p{2cm}|p{2cm}|p{2cm}|}
    \hline
    Antenna Length  & Resonance Freq.   & Transmitted Power ($P_t$)  & $A_e/\lambda^2$ \\
     $\mu m$& THz &$pW$&$\mu m^2$\\
    \hline
    \hline
    85&0.85& 1.65 &0.34\\
    \hline
    120&0.65& 2.93 &0.34\\
    \hline
    130&0.6& 3.26 &0.318\\
    \hline
    140&0.55& 3.67 &0.47\\
    \hline
    170&0.48& 4.89 &0.49\\
    \hline
    
  \end{tabular}
  \end{center}
\end{table}
     
\begin{figure}[htp]
    \centering
\includegraphics[width=\textwidth]{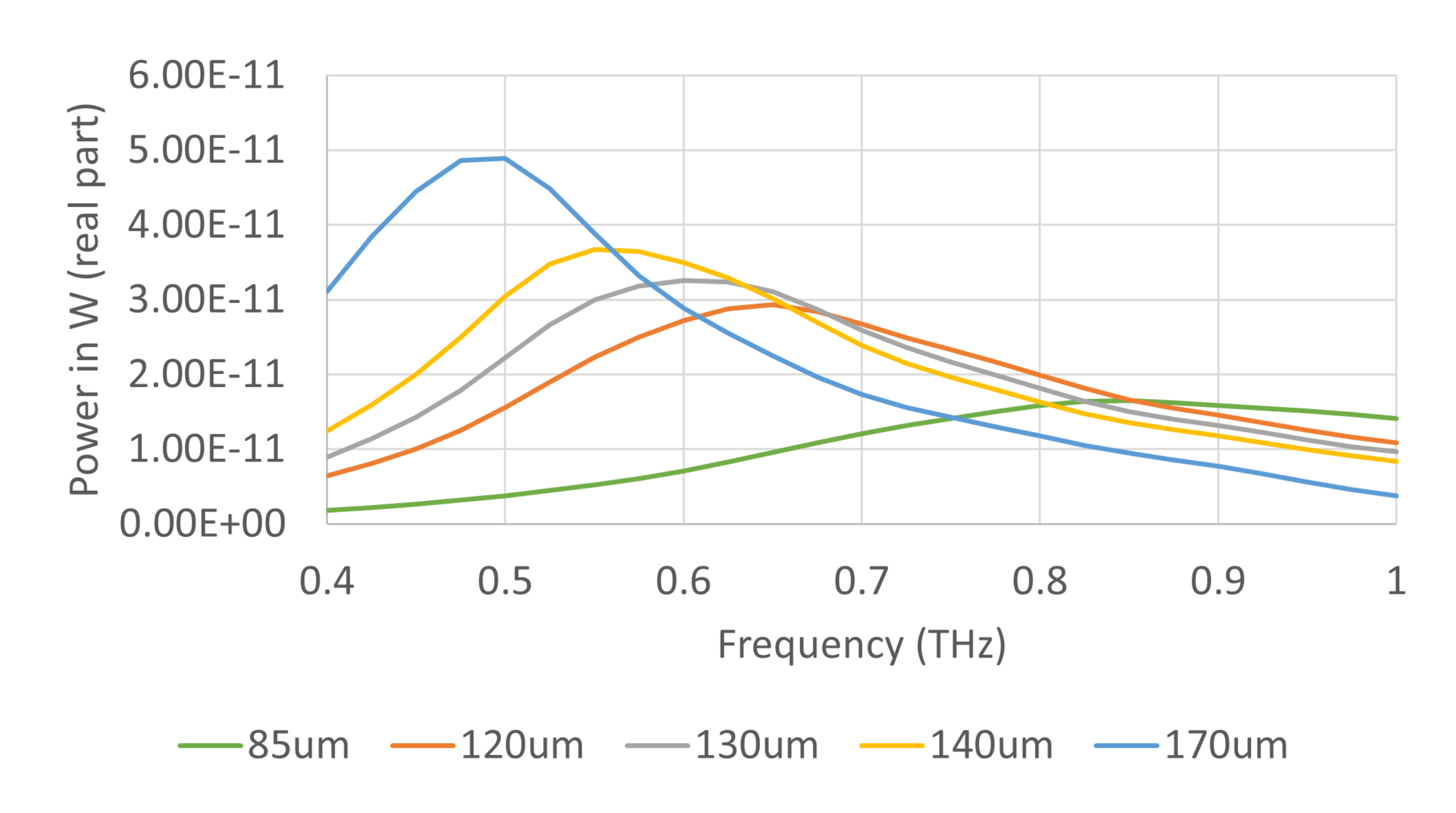}
\caption{FEM simulations of dipole antennas with varying the antenna arm’s length. As the antenna’s arms shrink, the effective aperture area of the antenna diminishes leading to a smaller power absorption amplitude at resonance.
}
    \label{figS2}
\end{figure}

\newpage
\begin{figure}[htp]
    \centering
\includegraphics[width=\textwidth]{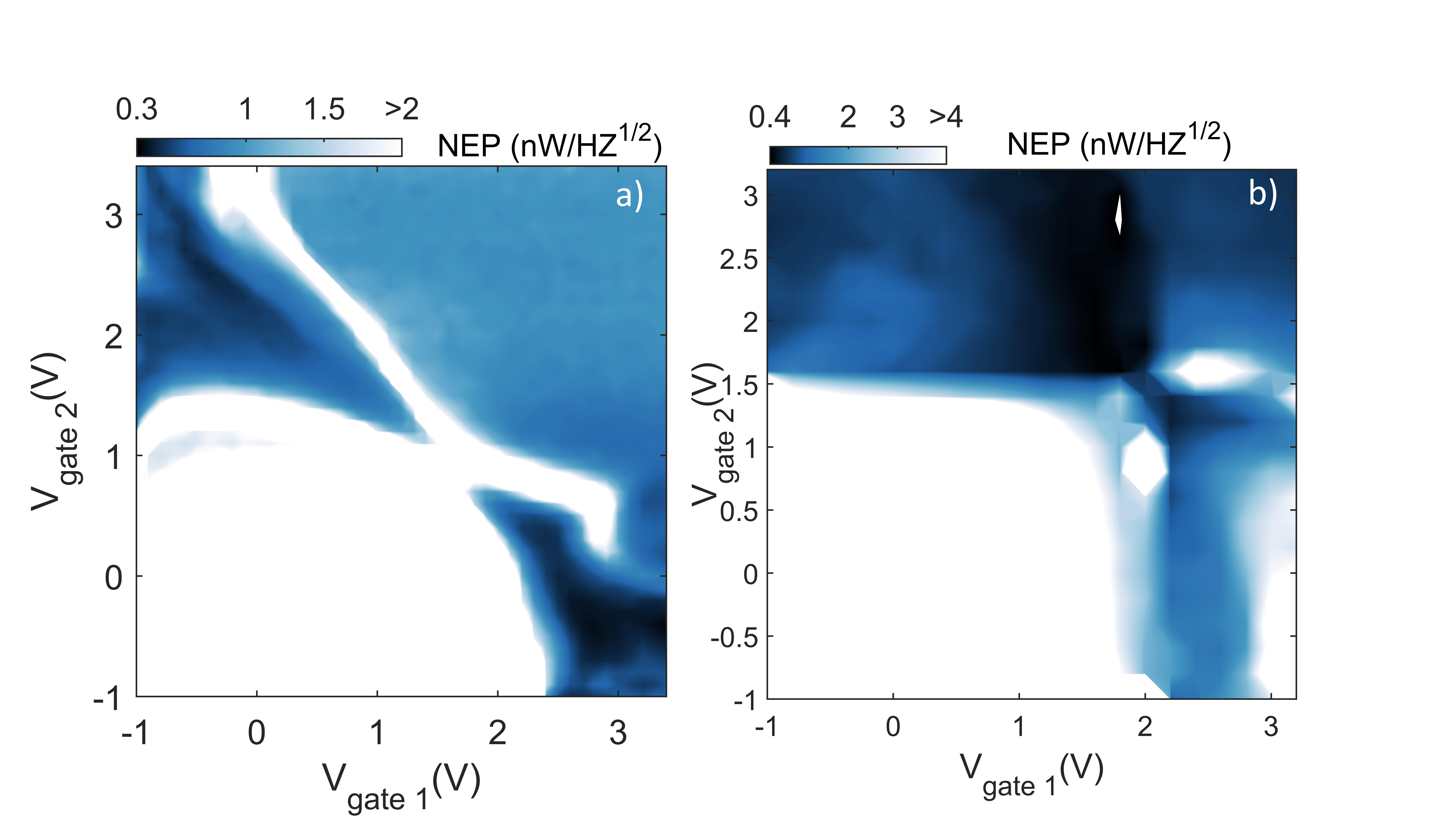}
\caption{Noise equivalent power map as a function of the gate's voltage for a) the patch antenna and b) the dipole antenna obtained from the measurements of $\rm I\!R_{THz}$ and noise voltage from Figure 2 and 3 in the main text.
}
    \label{figS1}
\end{figure}
\newpage
\begin{figure}[htp]
    \centering
\includegraphics[width=\textwidth]{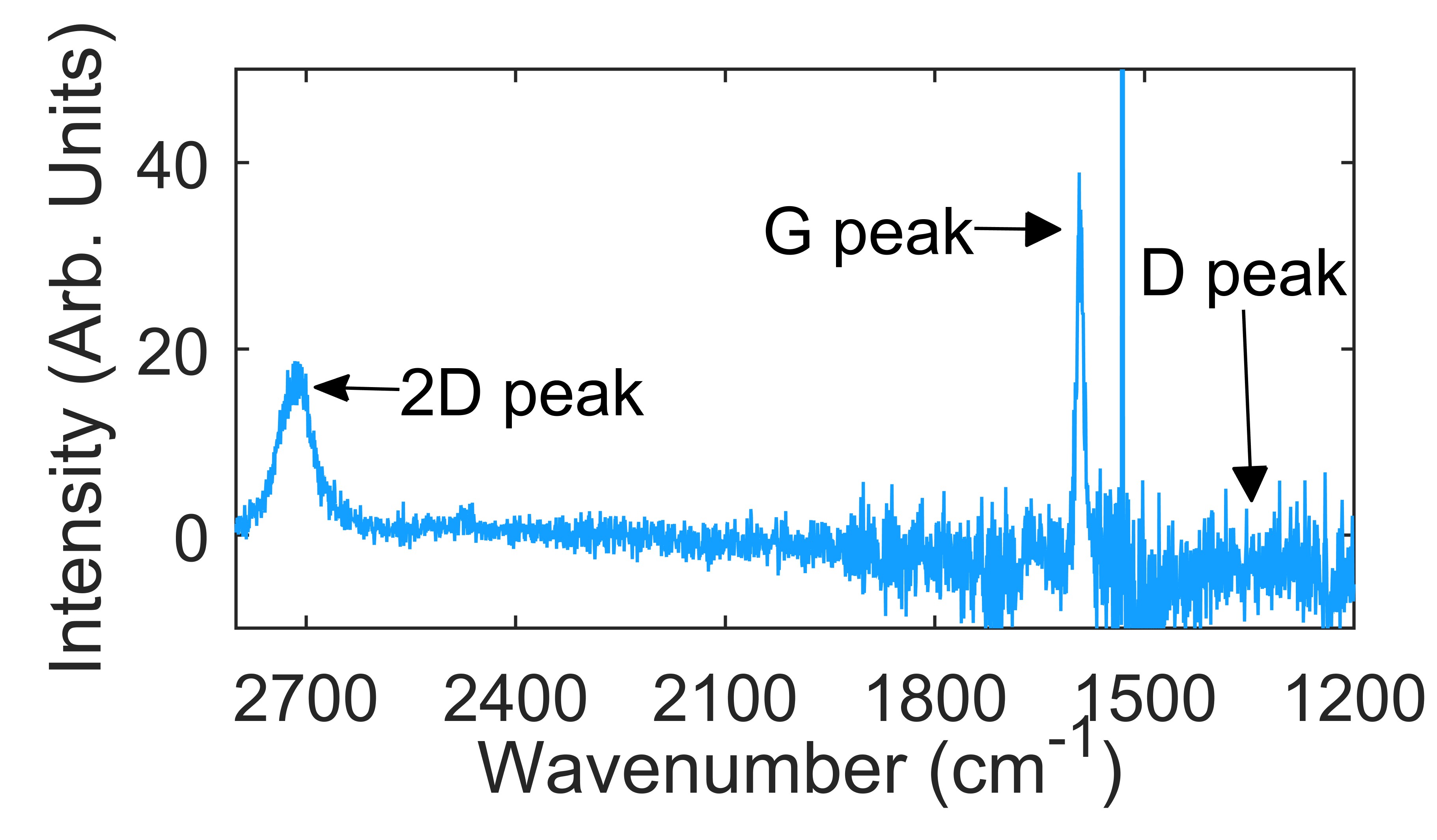}
\caption{Raman spectra from the quasi-free standing BLG with a broadened 2D peak due to the addition of an EG layer upon release of the buffer layer. The SiC substrate contributions were subtracted.
}
    \label{figS1}
\end{figure}
\newpage

\begin{figure}[htp]
    \centering
\includegraphics[width=\textwidth]{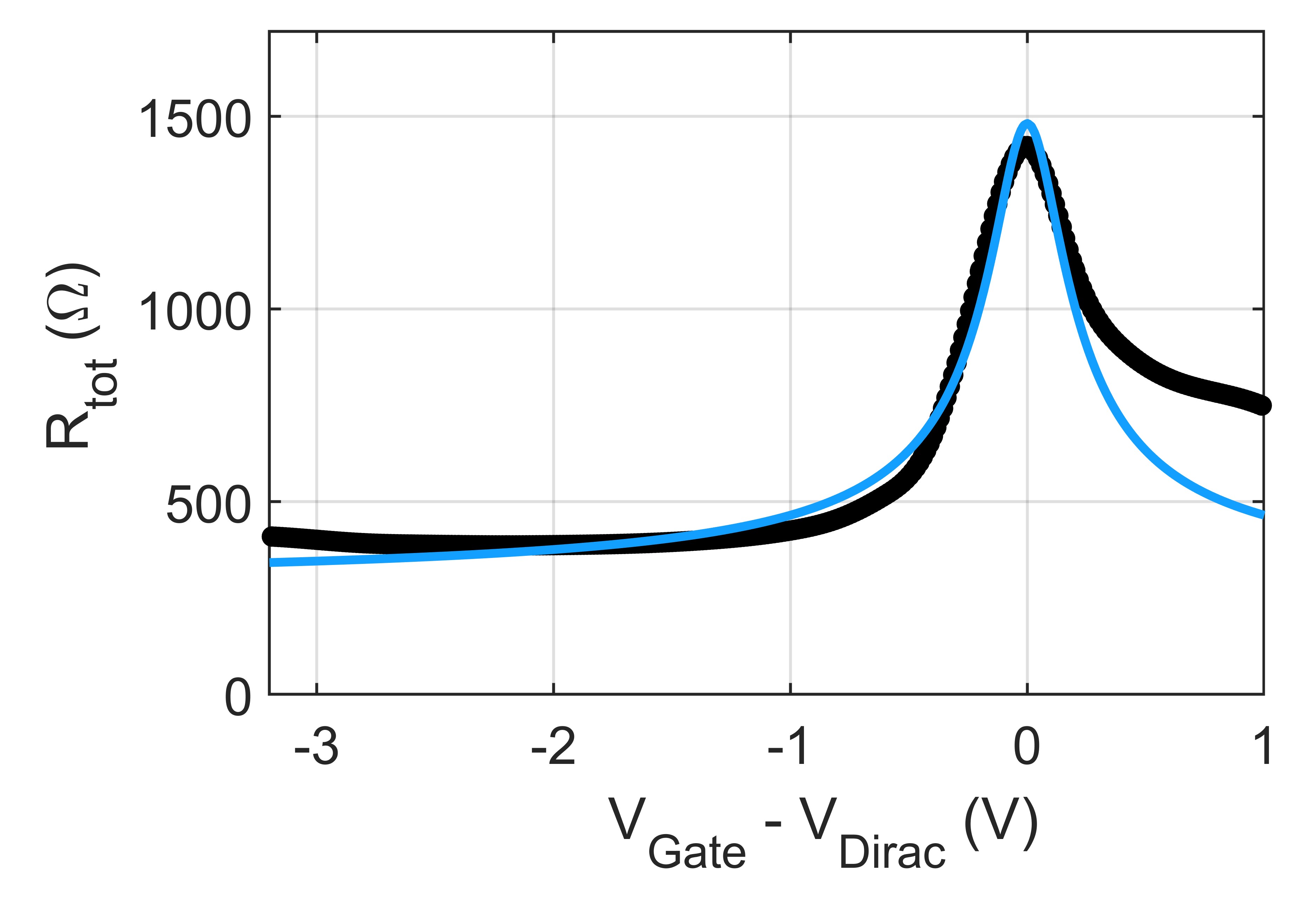}
\caption{Variation of total resistance ($R_{tot}$) as a function of gate voltage deviation from the Dirac point ($V_{gate} - V_{Dirac}$), alongside fitting results from Kim et al. (2009)\cite{kim2009realization}. The graphene channel, initially p-type, requires a gate voltage of approximately $V_{Dirac} = 2.2,V$ to align with the charge neutrality point (CNP). Through fitting the channel resistance curve, we extracted the carrier mobility in the gated graphene channel as $\mu = 450,cm^2/(V\cdot s)$ and the intrinsic carrier density as $n_0 \approx 3.0 \times 10^{12},cm^{-2}$. The contact resistance is estimated to be about $R_C \approx 200,\Omega$, using the fit function $R = R_C + \frac{N_{sq}}{1/n^2 e \mu}$, where $n = \sqrt{n_0^2 + n_{gate}^2(V_{gate})}$. Discrepancies between the fitted and experimental data are partially attributed to the mobile charge within the $SiO_2$. The observed asymmetry between electron and hole branches around the CNP is likely due to differences in their scattering cross-section near charged impurities. The degradation of mobility in the gated graphene channel, compared to as-grown bilayer graphene, is potentially caused by the dry etching process, oxide layer formation, and other fabrication steps.
}
    \label{figS2}
\end{figure}
     
\begin{figure}[htp]
    \centering
\includegraphics[width=1\textwidth]{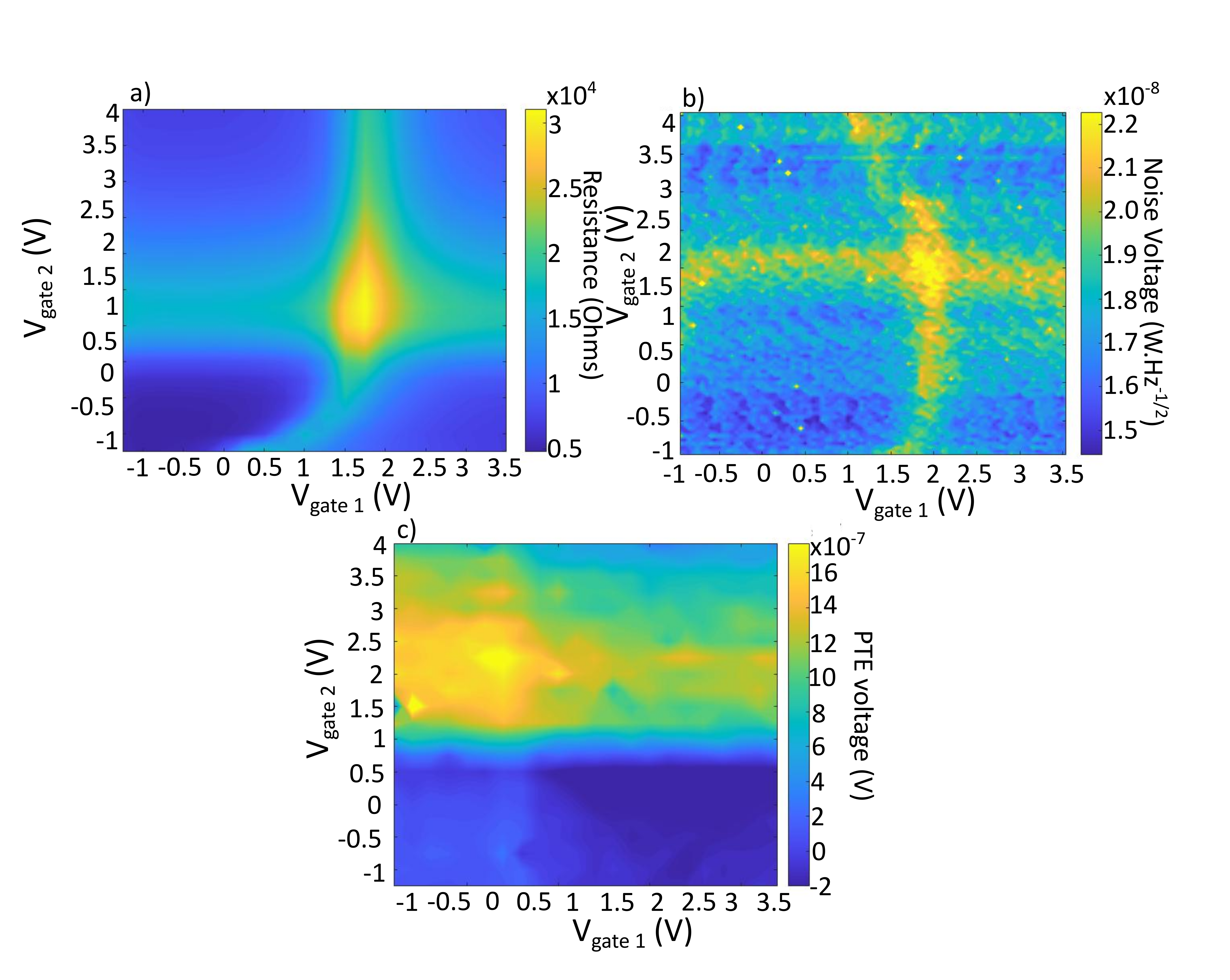}
\caption{Electro-optical characterization of a control dipole antenna PTE detector designed to have a maximum responsivity at 2.5THz. The control dipole spectra is displayed on Fig.4 of the main text: a) Resistance map as a function of the voltage applied to two top gates. b) Noise voltage map as a function of the gate's voltage with no terahertz excitation. c) THz PTE voltage map versus the gate’s voltage bias at 605 GHz.
}
    \label{figS2}
\end{figure}
     
\begin{figure}[htp]
    \centering
\includegraphics[width=0.4\textwidth,height=\textheight,keepaspectratio]{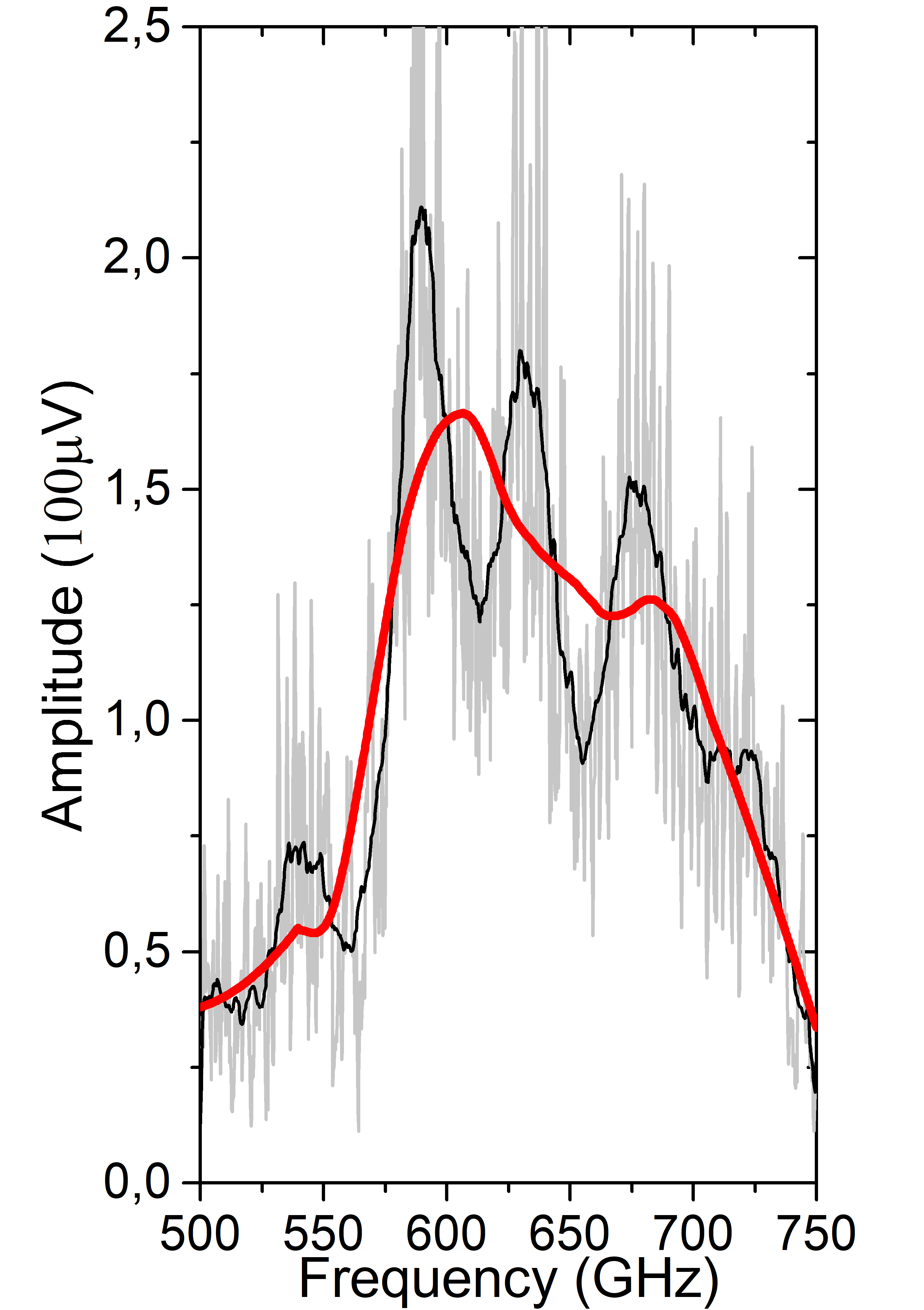}
\caption{Patch Antenna Spectrum showing the two steps of frequency filtering (photo-thermal voltage readout in function of the frequency). The raw spectrum is displayed in light grey and shows of a rapid Fabry-Perot oscillations (f=5GHz) which correspond to the total internal reflections of the 10mm thick Si lens. The first level of filtering reveals slow oscillations (f=45 GHz) whose origin is possibly from plasmonic oscillations on the read-out circuitry. The filtered spectrum is displayed in red.  
}
\label{figS7}
\end{figure}
      
\begin{figure}[htp]
    \centering
\includegraphics[width=0.5\textwidth,height=\textheight,keepaspectratio]{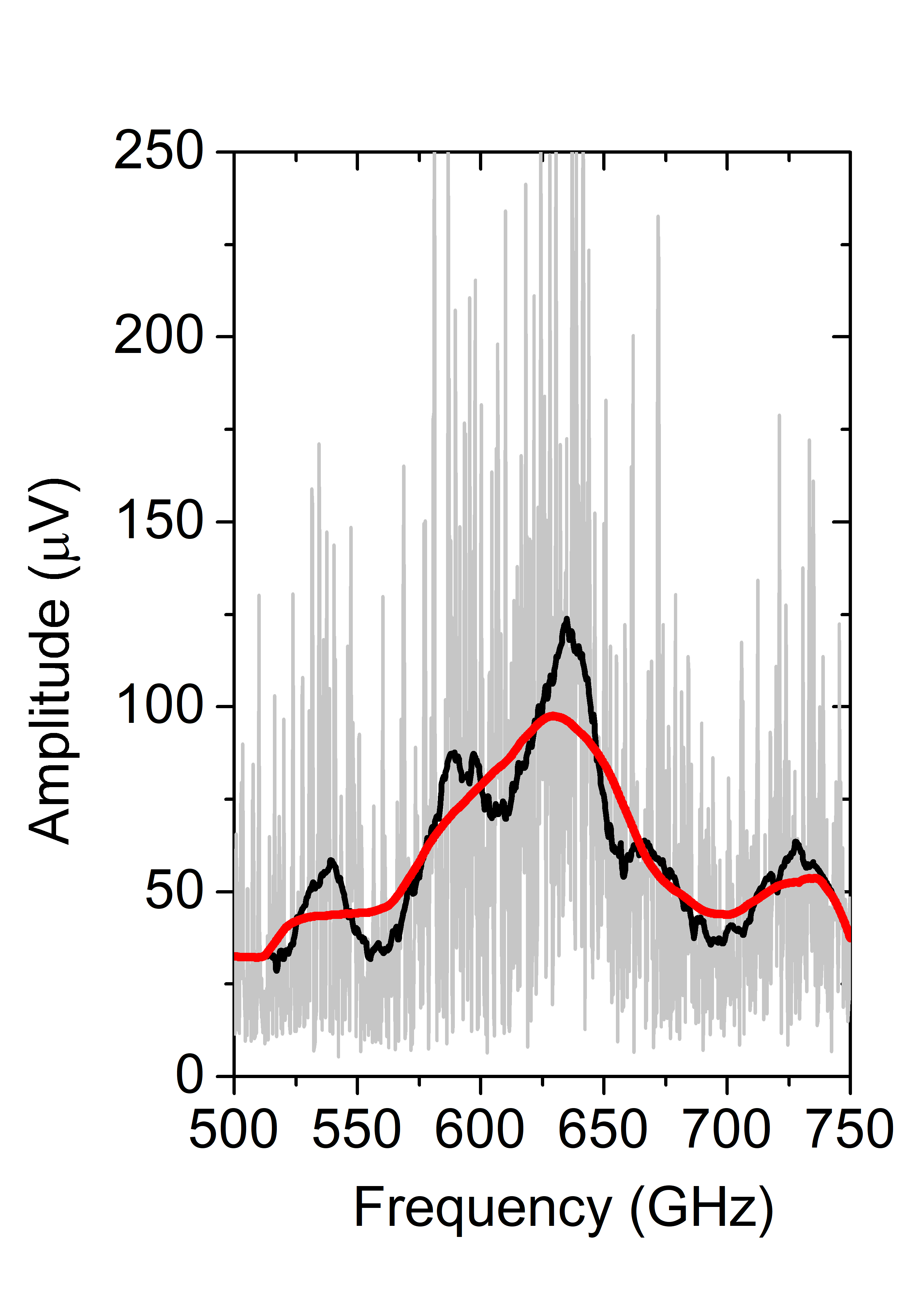}
\caption{Dipole Antenna Spectrum: photo-thermal voltage readout in function of the frequency for an optimised set of gate voltage ($V_{gate,1}$, $V_{gate,2})$. We used the same filtering steps as for the patch (Figure \ref{figS7})and dipole antenna spectra.}
\label{figS8}
\end{figure}
\cleardoublepage
\subsection{Responsivity Spectra at 5K}
\begin{figure}[htp]
    \centering
\includegraphics[width=0.8\textwidth]{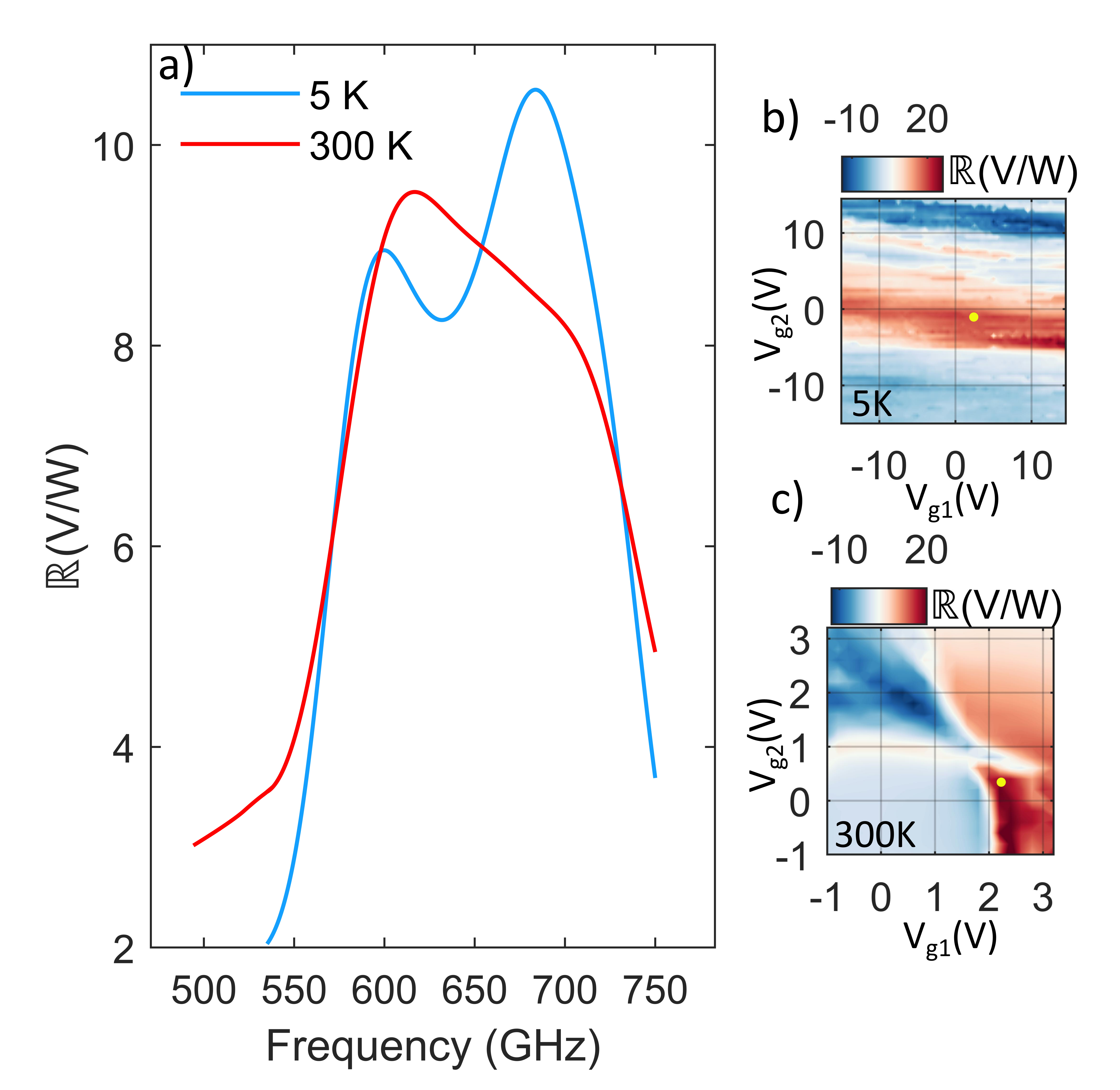}
\caption{A patch antenna BLG detector response spectrum (a) is recorded at room temperature and at 5K. b) and c) display the photo-voltage maps of the device taken at 5K and 300K respectively. The yellow bullets show at which gate voltage sets the spectra were recorded.
}
    \label{fige4}
\end{figure}

The responsivity enhancement of the antenna seems to be independent to the operating temperature. As the detector is cooled down to 5K, we observe a similar enhancement in the $\rm I\!R_{THz}$ as at room temperature but with them maximum absorption shifted to higher frequencies ($\Delta f=67\,GHz$). $\rm I\!R_{THz}$ and spectra at 300K and 5K are measured at zero bias voltage, for a set of gate voltage indicated by the yellow bullet in the insets. A change of the Fermi energy at 5K of the BLG has a direct impact on the graphene’s complex conductivity, resulting in a decrease of the kinetic inductance of the BLG sheet. The matching condition ($X_{A}=-X_{gap}$) occurs at a different frequency. The drastically different photo-voltage response map taken at 5K (Figure \ref{fige4}b) could be attributed to the decrease at low temperature of the BLG thermal conductance $G_{th}$to the SiC substrate, affecting the thermal dissipation of hot electrons in the channel\cite{pop2012thermal}. The length scale for heat flow to the contacts is the thermal healing length $l_{H}\simeq\sqrt{\kappa W_{A}h/G_{th}}$, where $W_{A}$ is the graphene channel width, $h$ the spacing between layers and $\kappa$ the thermal conductivity. A decrease in $G_{th}$ would increase the healing length $l_{H}$, enhancing cooling at the graphene/metal contact interfaces. An electronic temperature gradient, generated by the impinging THz field, would occur only near the graphene/metal p-n junctions. In this situation, the recorded PTE voltage at 5K only originates from the graphene/metal contact underneath the gate voltage 2 (which shows a tunability of the chemical potential, y-axis in Figure \ref{fig4}b ) while the second graphene/metal contact is grounded for the experiment, therefore not displaying any PTE responses. The splitting of the resonance frequency at 5K originating from the shift in the kinetic inductance in the BLG could be further exploited to generate frequency-tunable sensing platforms.

\cleardoublepage
\bibliography{supp}